\documentstyle[preprint,aps,epsfig,floats,psfig,graphicx,rotating]{revtex}
\tightenlines
\begin{document}
\pagestyle{empty}
%\title{
{\normalsize\hfill{\vspace*{-1.5cm}
         \vbox{\hbox{CERN-TH/2002-296}
         \hbox{YITP-SB-02-53}
         \hbox{IFIC/02-50}
%          \hbox{hep-ph/yymmddd}
}}}\\\vskip 1.5cm
\begin{center}
{\large \bf Neutrino Masses and Mixing: Where We Stand \\
\vskip 0.1cm
and Where We are Going
\footnote{
Review talk given at the 10th International Conference on Supersymmetry
and Unification of Fundamental Interactions, SUSY02 (June 17-23, 2002, 
DESY,  Hamburg).}
}\\
\vskip 0.4cm
{\large M. C. Gonzalez-Garcia} 
\vskip 0.2cm
{\sl\small Theory Division, CERN, CH-1211, Geneva 23, Switzerland,\\
  Y.I.T.P., SUNY at Stony Brook, Stony Brook,NY 11794-3840\\
and IFIC, Universitat de Val\`encia -- C.S.I.C., Apt 22085, 46071
  Val\`encia, Spain}\\
\end{center}
\vspace*{-0.5cm}
\begin{abstract}
In this talk I review our present knowledge on neutrino masses
and mixing as well as the expectations from near 
future experiments.
\end{abstract} 
\vspace*{-0.1cm}
\section{Introduction: What We Want to Learn and How}
\label{osci}
\vspace*{-0.2cm}
Our present understanding of neutrino masses and mixing
comes from the interpretation of data on solar and atmospheric 
neutrinos as well as the results from the LSND experiment in terms 
of neutrino oscillations~\cite{review}.

If neutrinos have masses, flavour is mixed in the  CC interactions
of the leptons, and a leptonic mixing matrix will
appear analogous to the CKM~\cite{ckm} matrix for the quarks.
The possibility of arbitrary mixing between two massive neutrino 
states was first introduced in Ref.~\cite{MNS}.
The discussion of leptonic mixing in generic models is complicated
by two factors. First the number 
massive neutrinos ($n$) is unknown, since there are
no constraints on the number of right-handed, SM-singlet,
neutrinos ($m=n-3$). Second, since neutrinos carry neither 
color nor electromagnetic charge, they could be Majorana
fermions. In general the mixing matrix in the CC current is a
$3\times n$ matrix which contains $\frac{n(n-1)}{2}$ mixing angles
and $\frac{(n-1)(n-2)}{2}$ phases if neutrinos are Dirac particles
and $n-1$ additional phases if neutrinos are Majorana particles.

The presence of the leptonic mixing, allows for flavour oscillations
of the neutrinos~\cite{pontecorvo}. A neutrino of energy $E$ produced in a 
CC interaction with a charged lepton $l_\alpha$ can be detected 
via a CC interaction with a charged lepton $l_\beta$ with a probability
\begin{equation}
{ P_{\alpha\beta}} =\delta_{\alpha\beta}-4\sum_{i< j}^n
{ \mbox{Re}[U_{\alpha i}U^*_{\beta i} U^*_{\alpha j} U_{\beta j}]}
\sin^2x_{ij} +2
\sum_{i<j}^n
{\mbox{Im}[U_{\alpha i}U^*_{\beta i} U^*_{\alpha j} U_{\beta j}]}
\sin^2\frac{x_{ij}}{2}
 \; ,  \label{pab}
\end{equation}
where in the convenient units 
$x_{ij}=1.27 \frac{\Delta m^2_{ij}}{\rm eV^2} \frac{L/E}{\rm m/{\rm MeV}}$,
with  $\Delta m^2_{ij} \equiv m_i^2-m_j^2$.
$L=t$ is the distance between the production point of
$\nu_\alpha$ and the detection point of $\nu_\beta$.
The first term in  Eq.~(\ref{pab}) is CP conserving while the second one
is CP violating and has opposite sign for $\nu$ and $\bar\nu$.

The transition probability  [Eq.~(\ref{pab})] presents an oscillatory 
behaviour, with oscillation lengths  
$L_{0,ij}^{\rm osc}=\frac{4 \pi E}{\Delta m_{ij}^2}$
and amplitude that is proportional to elements in the mixing matrix. 
From Eq.~(\ref{pab}) we find that neutrino oscillations are only sensitive
to mass squared differences. Also, the Majorana phases cancel out and only the
Dirac phase is observable in the CP violating term. 
In order to be sensitive to a given value of 
$\Delta m^2_{ij}$, an experiment has to be set up with $E/L\approx 
\Delta m^2_{ij}$ ($L\sim L_{0,ij}^{\rm osc}$). 

For a two-neutrino case, the mixing matrix depends on a single parameter,
there is a single mass-squared difference $\Delta m^2$ and there is 
no Dirac CP phase.
Then $P_{\alpha\beta}$ of Eq.~(\ref{pab}) takes the well known form  
\begin{equation}
P_{\alpha\beta}=\delta_{\alpha\beta}- (2\delta_{\alpha\beta}-1) \sin^22\theta 
\sin^2x \;.
\label{ptwo}
\end{equation} 
The full physical parameter space is covered with $\Delta m^2\geq 0$ 
and $0\leq\theta\leq\frac{\pi}{2}$ (or, alternatively,
$0\leq\theta\leq\frac{\pi}{4}$ and either sign for $\Delta m^2$).
Changing the sign of the mass difference, $\Delta m^2\to-\Delta m^2$, and
changing the octant of the mixing angle, $\theta\to\frac{\pi}{2}-\theta$,
amounts to redefining the mass eigenstates, $\nu_1\leftrightarrow\nu_2$:
$P_{\alpha\beta}$ must be invariant under such transformation. 
Eq.~({\ref{ptwo}) reveals, however, that $P_{\alpha\beta}$ is actually
invariant under each of these transformations separately. This situation 
implies that there is a two-fold discrete ambiguity in the interpretation
of $P_{\alpha\beta}$ in terms of two-neutrino mixing: the two different sets of
physical parameters, ($\Delta m^2, \theta$) and ($\Delta m^2, \frac{\pi}{2}
-\theta$), give the same transition probability in vacuum. One cannot tell from
a measurement of, say, $P_{e\mu}$ in vacuum whether the larger component of 
$\nu_e$ resides in the heavier or in the lighter neutrino mass eigenstate. 

This symmetry is lost when neutrinos travel through regions of dense matter. 
In this case, they can undergo forward scattering with
the particles in the medium. These interactions are, in general, 
flavour dependent and they can be included as a potential
term in the evolution equation of the flavour states. As a consequence 
the oscillation pattern is modified. Let us consider, for instance, 
oscillations $\nu_e\rightarrow \nu_\mu$ in a neutral medium
like the Sun or the Earth. For this system, the instantaneous
mixing angle in matter takes the form
\begin{equation} 
{\sin 2\theta_{m}}= 
\frac{{ \Delta{m}^2 \sin 2\theta }} 
{\sqrt{ ({\Delta{m}^2 \cos 2\theta } -A)^2 
+({ \Delta{m}^2 \sin 2\theta} )^2 }} 
\label{effmix}
\end{equation}
where $A=2 E V_{\rm CC}=2 \sqrt{2} E G_F N_e$ ($N_e$ is the electron number density
in the medium). Eq~(\ref{effmix}) shows an enhancement (reduction) 
of the mixing angle in matter for $\theta<\frac{\pi}{4}$ 
($\theta>\frac{\pi}{4}$)~\cite{msw}. Thus, matter effects allow to 
determine whether the larger component of 
$\nu_e$ resides in the lighter neutrino mass eigenstate.
As we will see this is the presently favoured scenario 
for solar neutrino oscillations. 
For mixing of three or more neutrinos, the oscillation probability,
even in vacuum, does not depend in general of $\sin^2 2\theta_{ij}$.

The experiments I will discuss in this talk give information 
on some $P_{\alpha\beta}$ which we interpret in terms of
neutrino masses and mixing. It was common
practice to make this interpretation in the two-neutrino
framework and translate the  constraints on $P_{\alpha\beta}$ 
into allowed or excluded regions in the plane 
($\Delta m^2,\; \sin^22\theta$). 
However, as we have seen once matter effects are important, or mixing among 
more than two neutrinos is considered, the covering of the full parameter
space requires the use of a single-valued function of the mixing angle
such as $\sin^2\theta$ or $\tan^2\theta$~\cite{dark}. 
\section{Solar Neutrinos}
\vspace*{-0.2cm}
\subsection{The Evidence}
\vspace*{-0.2cm}
The sun is a source of $\nu_e's$ which are produced in the different
nuclear reactions taking place in its interior. Along this talk I will 
use the $\nu_e$ fluxes from Bahcall--Pinsonneault 
calculations~\cite{bp00} which I refer to as the solar standard model (SSM).
These neutrinos have been detected at the Earth by seven
experiments which use different detection 
techniques~\cite{chlorine,gallex,gno,sage,kamsun,sksolar,sno01,sno02}
Due to the different energy threshold and the different detection
reactions, the experiments  are sensitive to different parts of the 
solar neutrino spectrum and to the flavour composition of the beam.
In table~\ref{tab:solarexp} I show the different experiments and detection
reactions with their energy threshold  as well as
their latest results on the total event rates as compared to 
the SSM prediction. 
\begin{table*}
\begin{center}
\begin{tabular}
{llccc}
 Experiment & { Detection}  & { Flavour} 
& { $ E_{\rm th}$ (MeV)} & { $\frac{\rm Data}{\rm BP00}$} \\[+0.1cm]
\hline 
{ Homestake} &  { $^{37}$Cl$(\nu,e^-)^{37}$Ar} & { $\nu_e$} 
& { $E_\nu> 0.81$} & { $0.34\pm 0.03$}\\[+0.1cm]  
{ Sage +}  & & &  &  \\ [-0.2cm]
& { $^{71}$Ga$(\nu,e^-)^{71}$Ge} & { $\nu_e$}
& { $E_\nu> 0.23$} & { $0.56\pm 0.04$}\\ [-0.2cm] 
{ Gallex+GNO} & & &  &  \\[+0.2cm]
&   &      { $\nu_e$, $\nu_{\mu/\tau}$} & &  \\[-0.2cm]
{ Kam     
$\Rightarrow$ SK} &  
{ ES $\;\nu_x e^- \rightarrow \nu_x e^-$} &  & 
{ $E_e > 5$} & { $0.46\pm 0.02$} \\[-0.2cm]
&            & {\footnotesize
$\left(\frac{\sigma_{\mu\tau}}{\sigma_e}\simeq \frac{1}{6}\right)$} 
&            &        \\[+0.1cm]
{ SNO} & { CC $\;\nu_e d \rightarrow p p e^-$} &
${\nu_e}$ &{ $T_e>5$}  & { $0.35\pm 0.02$} \\
  & { NC $\;\nu_x d \rightarrow \nu_x d$} &
{ $\nu_e$, $\nu_{\mu/\tau}$} 
&{ $T_\gamma >5$}  & { $1.01\pm 0.12$} \\
  & { ES $\;\nu_x e^- \rightarrow \nu_x e^-$} &
{ $\nu_e$, $\nu_{\mu/\tau}$} 
 &{ $T_e >5$}  & { $0.47\pm 0.05$} \\
\end{tabular}
\end{center}
\label{tab:solarexp}
\caption{Event rates observed at solar neutrino experiments compared to the
SSM predictions (the errors do not include the theoretical uncertainties).
For SNO, the quoted rates are obtained under the  hypothesis of 
undistorted $^8$B spectrum.}
\end{table*}

We can make the following statements: 
\begin{itemize} 
\item Before the NC measurement at SNO all experiments observed a flux 
that was smaller than the SSM predictions, $\Phi^{\rm obs}/\Phi^{\rm
  SSM}\sim0.3-0.6$. 
\item The deficit is not the same for the various experiments, 
which indicates that the effect is energy dependent.
\item SNO has observed an event rate different in the different reactions.
In particular in NC SNO observed no deficit as compared to the SSM.
\end{itemize}
The first two statements constitute the solar neutrino problem. The last one,
has provided us in the last year with evidence of 
flavour conversion of solar neutrinos independent of the solar model.

Both SK and SNO measure the high energy $^8$B neutrinos. 
Schematically, in presence of flavour conversion the observed fluxes
in the different reactions are
\begin{eqnarray}
\Phi^{\rm CC}&=&\Phi_e, \nonumber \\
\Phi^{\rm ES}&=&\Phi_e\,+\, r\,\Phi_{\mu\tau}, \\
\Phi^{\rm NC}&=&\Phi_e+ \Phi_{\mu\tau}, \nonumber 
\end{eqnarray}
where 
$r\equiv \sigma_{\mu}/\sigma_{e}\simeq 0.15$ 
is the ratio of the the $\nu_e - e$ and $\nu_{\mu} - e$ elastic 
scattering cross-sections. The flux
$\Phi_{\mu\tau}$ of active no-electron neutrinos is zero in the SSM.
Thus, in the absence of flavour conversion, the three observed 
rates should be equal.  The first reported SNO CC~\cite{sno01} result 
compared with the ES rate from SK~\cite{sksolar} showed that the hypothesis of
no flavour conversion was excluded at $\sim 3\sigma$. 
Finally, with the NC measurement at SNO~\cite{sno02} one finds that 
\begin{equation}
\Phi_{\mu\tau}=
(3.41\pm 0.45 ^{+0.48}_{-0.45}) \times 10^6\ {\rm cm}^{-2} {\rm s}^{-1}.
\label{snosk}
\end{equation}
This result provides evidence for neutrino flavor transition (from
$\nu_e$ to $\nu_{\mu,\tau}$) at the level of $5.3\sigma$. This
evidence is independent of the solar model.

SK and SNO also gave information on the time and energy
variation of their signals:\\
-- The recoil electron energy spectrum measured at SK and the 
effective energy spectrum of SNO show no evidence of energy dependence
beyond the expected in the SSM. \\
--The Day/Night variation which measures
the effect of the Earth Matter in the neutrino propagation.
Both SK and SNO finds a few more events at night than 
during the day but with very little statistical significance. 

In order to combine both the Day--Night information and the
spectral data SK has also presented separately the measured recoil
energy spectrum during different zenith angle bins 
(a total of 44 data points) 
while  SNO has presented their results as day and night spectrum
(34 data points). 

\subsection{The Interpretation}
The most generic and popular explanation to this observation is 
in terms of neutrino masses and mixing leading to oscillations 
of $\nu_e$ into an active ($\nu_\mu$ and/or  $\nu_\tau$) or a sterile 
($\nu_s$) neutrino.  Several global analyses of the solar neutrino data
have appeared in the literature after the latest SNO results~\cite{solana}.
In Fig.~\ref{solarosc} I show the results of a global 
analysis~\cite{oursolar} of the latest solar neutrino data in 
terms of oscillation parameters. 

All analysis  find that active oscillations are clearly favoured. 
LMA is the best fit and the only solution at $\sim$ 99\%CL. 
At 3$\sigma$ the allowed parameter  space  within LMA is in the first octant  
and there is an upper bound  of $\Delta m^2\lesssim 4\times 10^{-4}$ eV$^2$
~\cite{oursolar}
(the precise range of masses and mixing are slightly different for the
different analysis).
SMA is ruled out at $\sim 4\sigma$ as a consequence of the tension between
the low rate observed by SNO in CC and the flat spectrum observed by SK.  
Sterile oscillations are disfavoured at $\sim 5 \sigma$ 
due to the difference between the observed CC and NC event rates at
SNO. In reaching this conclusions both the more detailed information on the
day-night spectrum of SK and the new SNO results have played very 
important and complementary roles.
\begin{figure}[ht]
\begin{center}
\mbox{\epsfig{file=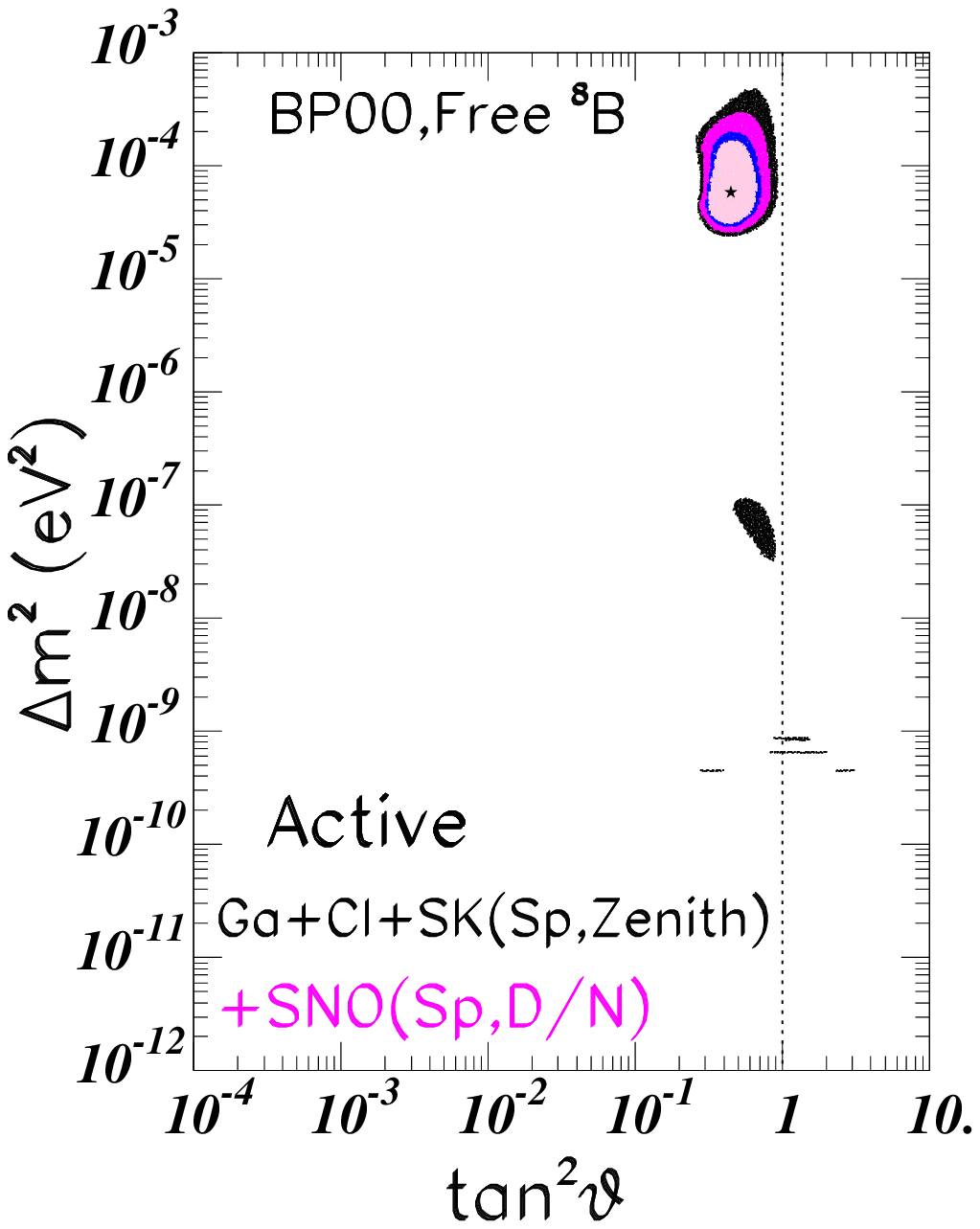,width=0.4\textwidth}} 
\mbox{\epsfig{file=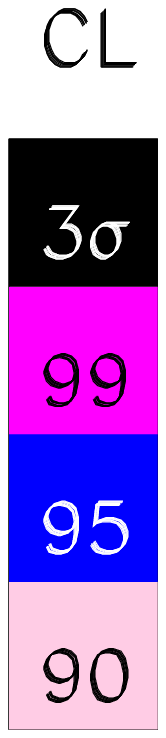,width=0.1\textwidth}} 
\end{center}
\caption{Allowed regions from the global fit for solar neutrino oscillations.}
\label{solarosc}
\end{figure}

\subsection{The Future: KamLAND and Borexino}
Our present understanding of the solar neutrino oscillation is 
being tested in the 
KamLAND experiment  which is currently in operation in the 
Kamioka mine in Japan. This
underground site is conveniently located at a distance of 150-210 km 
from several Japanese nuclear power stations. The measurement of the flux and 
energy spectrum of the $\bar\nu_e$'s emitted by these reactors will 
provide a test to the LMA solution of the solar neutrino anomaly~\cite{kland}.

After two or three  years of data taking, 
KamLAND should be capable of either excluding the entire LMA region or, 
not only establishing 
$\nu_{e}\leftrightarrow \nu_{\rm other}$ oscillations, but also 
measuring the LMA oscillation parameters with unprecedented 
precision~\cite{mura,kland2,sterile} provided that 
$\Delta m^2\leq {\rm few} 10^{-4}$ eV$^2$.
In Fig.~\ref{fig:kland} I show 
the expected precision in the construed oscillation 
parameters if KamLAND observes a signal corresponding to the 
expectations at the best fit point of the LMA region ~\cite{sterile}.
KamLAND is expected to announce their first results this year. 
For the purpose of illustration I also show in  Fig.~\ref{fig:kland}
the results of a simulation of the expected accuracy with $\sim$ 
three months of data ~\cite{carlos}.
\begin{figure}
\begin{center}
\includegraphics[scale=0.45]{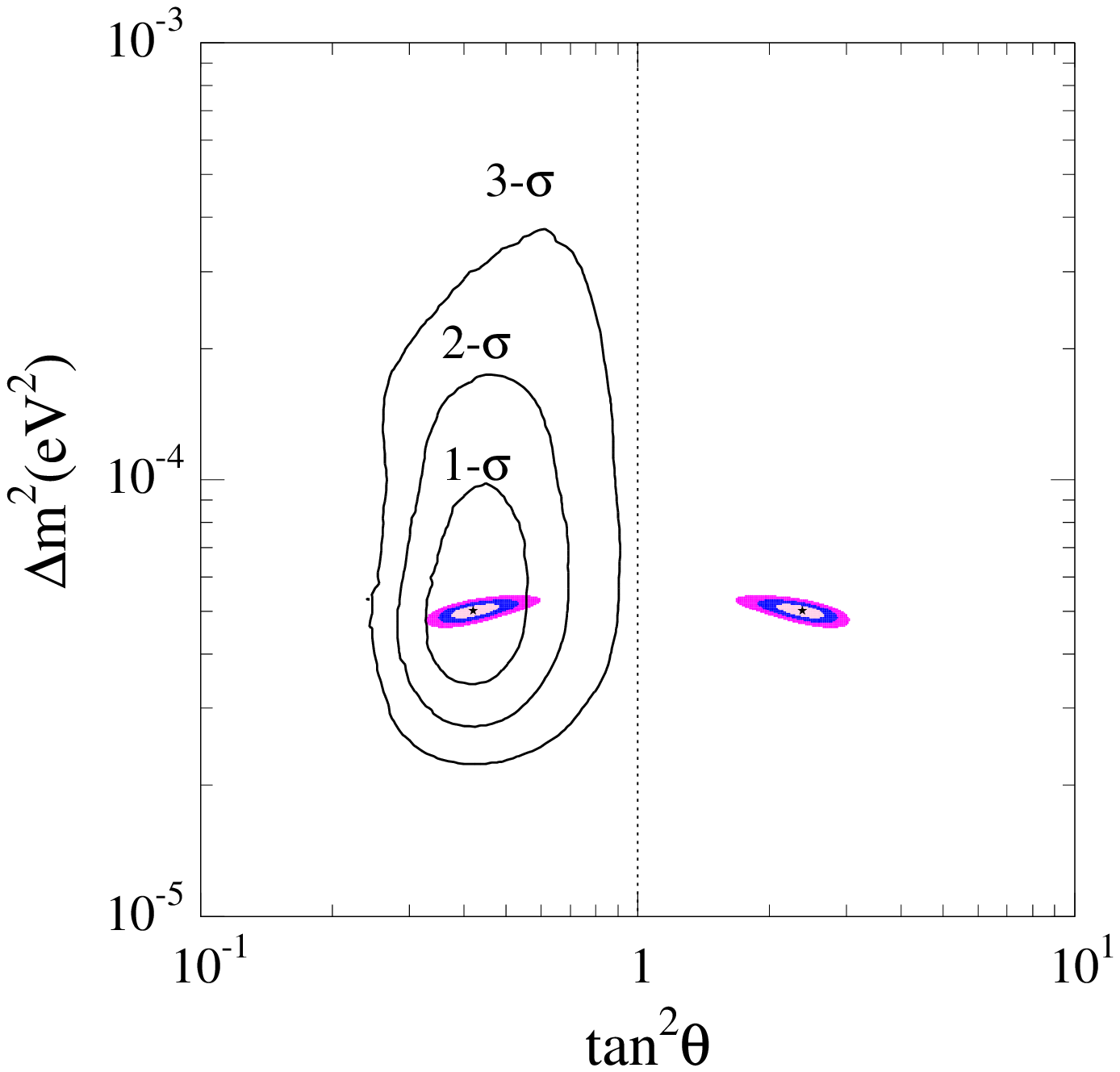}
\includegraphics[scale=0.43]{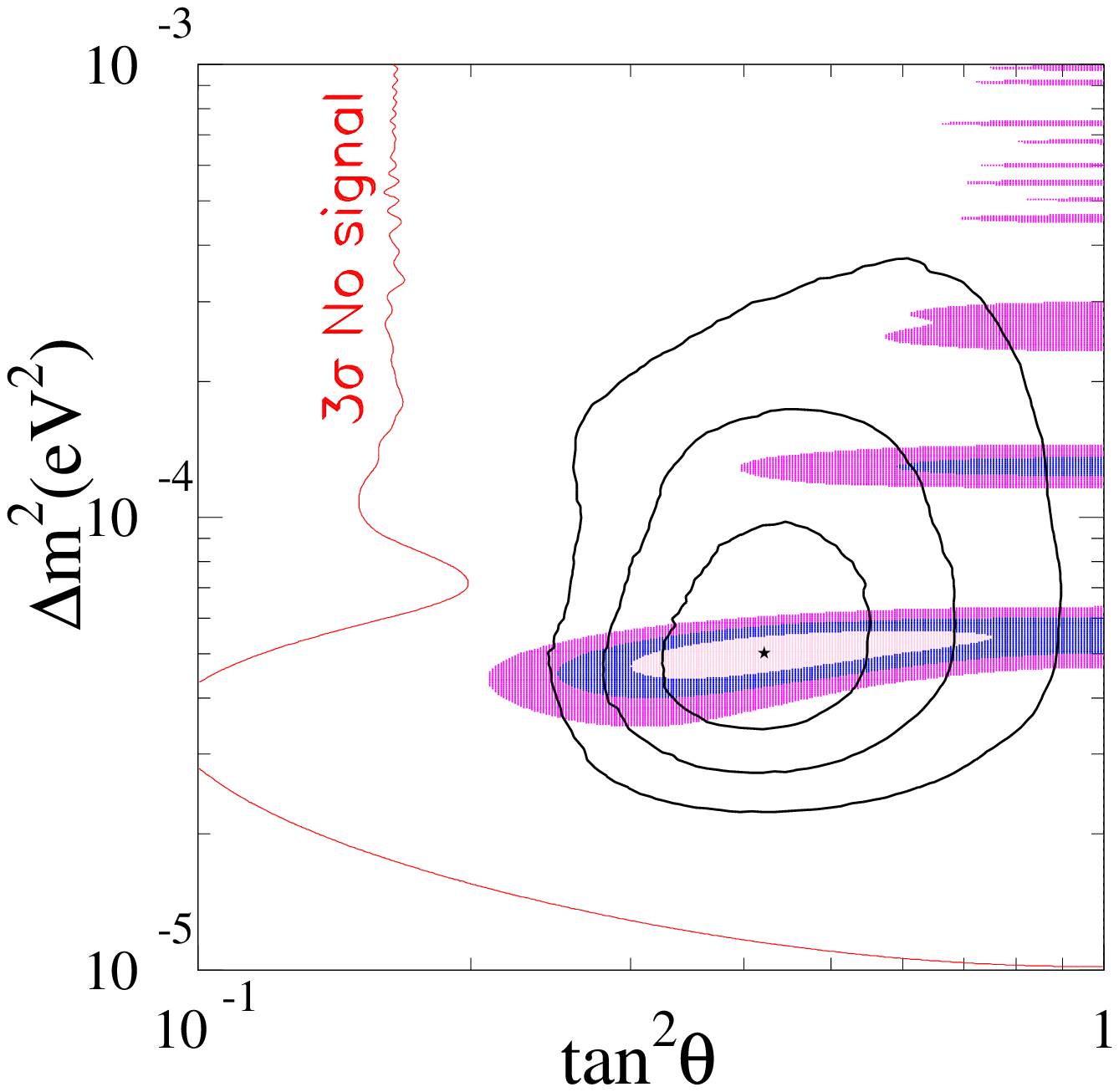}
\end{center}
\caption{{\sl Left}: 
Expected reconstructed regions (at 1, 2, and 3 $\sigma$ CL) 
if KamLAND observes a signal
corresponding to the present best fit point of the LMA region.
after three years of running. {\sl Right}: Same but after three months
of data taking (only the first octact in shown but the KamLAND
regions are symmetric around maximal mixing since matter
effects are negligible). The area to the right of the red curve will be ruled
out if no oscillation signal is observed.} 
\label{fig:kland}
\end{figure}

If LMA is confirmed, CP violation may be  observable at future 
long-baseline (LBL)
experiments. Also KamLAND will provide us, as data accumulate, with a firm 
determination of the corresponding oscillation parameters. 
With this at hand, the future solar $\nu$ experiments will be able to return
to their original goal of testing solar physics. 

If KamLAND does not confirm LMA, the next most relevant results will come from
Borexino ~\cite{borexino}. 
The Borexino experiment 
is designed to detect low-energy solar neutrinos in real-time through the 
observation of the ES process $\nu_a + e^- \to \nu_a + e^- $. The energy 
threshold for the recoil electrons is 250 keV. The largest contribution 
to their expected event rate is from neutrinos of the $^7$Be line.
Due to the lower energy threshold, Borexino is sensitive to matter effects in 
the Earth in the LOW region. And because $^7$Be neutrinos 
are almost monoenergetic, it is also very sensitive to seasonal 
variations associated with VAC oscillations. 
 
\section{Atmospheric and Reactor Neutrinos}
\subsection{The Evidence}
Atmospheric showers are initiated when primary cosmic rays hit the
Earth's atmosphere. Secondary mesons produced in this collision,
mostly pions and kaons, decay and give rise to electron and muon
neutrino and anti-neutrinos fluxes whose interactions are 
detected in underground detectors~\cite{imb,kamatm,soudan,skatm,macro}.
Atmospheric neutrinos can be detected in underground detectors by direct 
observation of their charged current interaction inside the detector.
These are the so called contained events. 
SK has divided their contained data sample into
sub-GeV events with visible energy below 1.2 GeV and multi-GeV above such
cutoff. On average, sub-GeV events arise from neutrinos of several hundreds of
MeV while multi-GeV events are originated by neutrinos with energies of the
order of several GeV. Higher energy muon neutrinos
and antineutrinos can also be detected indirectly by observing the muons
produced in their charged current interactions in the vicinity of the
detector. These are the so called upgoing muons. Should the muon 
stop inside the detector, it will be classified as a ``stopping'' muon,
(which arises from neutrinos of energies around ten GeV)
while if the muon track crosses the full detector the event is 
classified as a ``through-going'' muon which is originated by neutrinos
with energies of the order of hundred GeV.
\begin{figure}[ht]
\vspace*{-0.5cm}
\begin{center}
\mbox{\epsfig{file=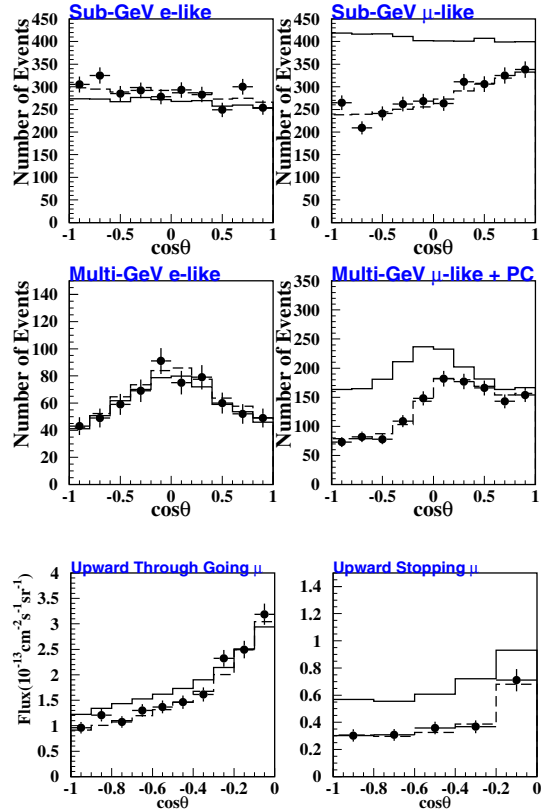,height=0.5\textheight}}
\end{center}
\caption{Zenith angle distribution of SuperKamiokande 1289 days
data samples. Dots, solid line and dashed line correspond
to data, MC with no oscillation and MC with best fit oscillation parameters,
respectively. 
Allowed parameters from the global fit of atmospheric neutrino
  data for $\nu_\mu\rightarrow \nu_\tau$ oscillations.}
\label{fig:skatm} 
\end{figure} 

At present the atmospheric neutrino anomaly (ANA) can be summarized in 
three observations (See Fig.~\ref{fig:skatm}): \\
-- There has been a long-standing deficit of about 60 \%  between the 
predicted and observed 
$\nu_\mu$$/\nu_e$ ratio of the contained events strengthened 
by the high statistics sample collected at the SK experiment. 
\\
-- The most important feature of the atmospheric neutrino
data at SK is that it exhibits a {\sl zenith-angle-dependent} deficit of 
muon neutrinos which indicates that the deficit is larger for muon neutrinos
coming from below the horizon which have traveled longer distances 
before reaching the detector. 
On the contrary, electron neutrinos behave
as expected in the SM. \\
-- The deficit for through-going muons is smaller that
for stopping muons, {\it i.e.} the deficit decreases as the neutrino 
energy grows.

\subsection{The Interpretation: Three-Neutrino Oscillations}
The most likely solution of the ANA involves neutrino
oscillations. At present the best solution from a global analysis of the
atmospheric neutrino data is $\nu_\mu\rightarrow \nu_\tau$ 
oscillations with oscillation parameters shown in Fig.~\ref{atmosc} 
\begin{figure}[ht]
\vspace*{-0.5cm}
\begin{center}
\includegraphics[scale=0.4]{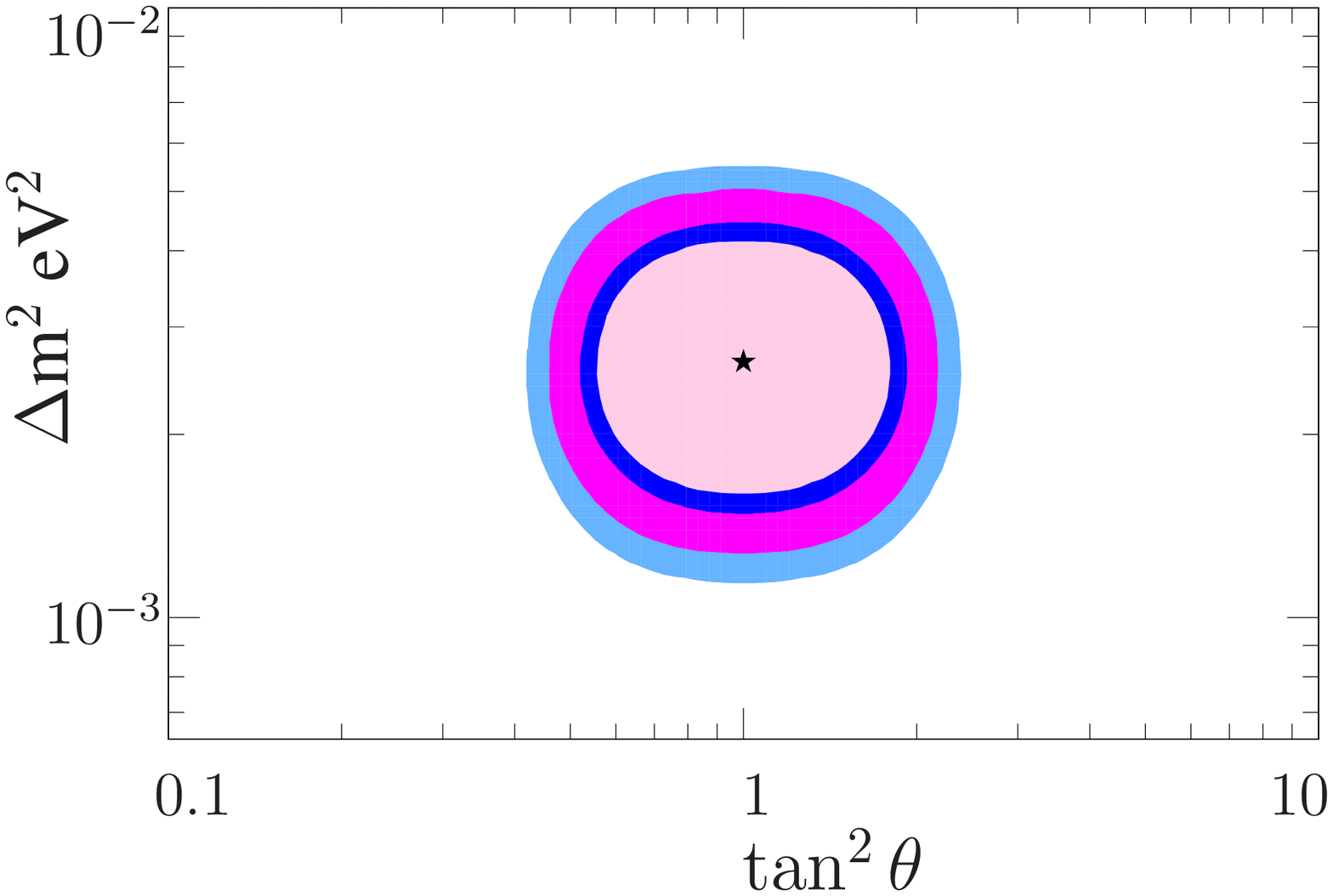}
\includegraphics[scale=0.5]{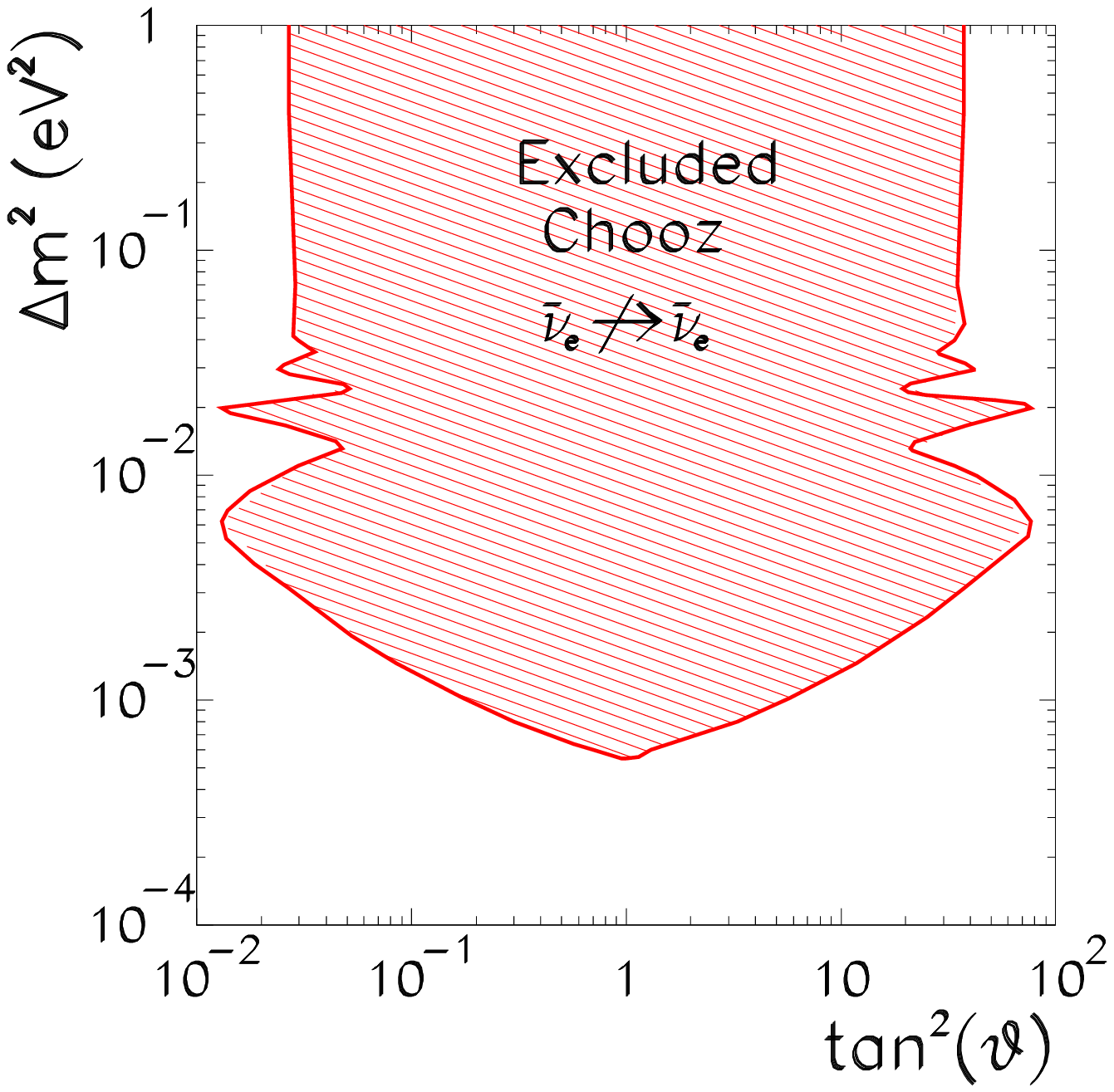}
\end{center}
\caption{{\sl Left:}
Allowed parameters from the global fit of atmospheric neutrino
  data for $\nu_\mu\rightarrow \nu_\tau$ oscillations.
{\sl Right:} 90\% CL excluded parameters by the CHOOZ experiment}
\label{atmosc} 
\end{figure} 

Oscillations into electron neutrinos are nowadays ruled out since
they cannot describe the measured angular dependence of muon-like
contained events. Moreover the most favoured range
of masses and mixings for this channel have been excluded by the 
negative results from the CHOOZ reactor experiment~\cite{chooz}.

The CHOOZ experiment~\cite{chooz} searched for disappearance of
$\bar{\nu}_e$ produced in a nuclear power station.  At
the detector, located at $L\simeq 1$~Km from the reactors, the
$\bar{\nu}_e$ reaction signature is the delayed coincidence between
the prompt ${\rm e^+}$ signal and the signal due to the neutron
capture in the Gd-loaded scintillator.  Their measured vs.\ expected
ratio, averaged over the neutrino energy spectrum is
\begin{equation}
    R = 1.01 \pm 2.8 \,\% ({\rm stat}) \pm 2.7 \,\% ({\rm syst}) \,.
    \label{eq:rchooz}
\end{equation}
Thus no evidence was found for a deficit of measured vs.\ expected
neutrino interactions, and they derive from the data exclusion plots
in the plane of the oscillation parameters in the simple two-neutrino 
oscillation scheme as shown~Fig.~\ref{atmosc}. 

Oscillations into sterile neutrinos are also disfavoured because
due to matter effects in the Earth they predict a flatter-than-observed  
angular dependence of the through-going muon data~\cite{skatm}.

The minimum joint description of atmospheric, solar and
reactor data requires 
that all three known neutrinos take part in the oscillations. 
The mixing
parameters are encoded in the $3 \times 3$ lepton mixing matrix 
which can be conveniently parametrized 
in the standard form 
\begin{equation}
U=\pmatrix{1&0&0 \cr 0& {c_{23}} & {s_{23}} \cr
0& -{s_{23}}& {c_{23}}\cr}\pmatrix{ 
{c_{13}} & 0 & {s_{13}}e^{i {\delta}}\cr
0&1&0\cr -{ s_{13}}e^{-i {\delta}} & 0  & 
{c_{13}}\cr}  \pmatrix{{c_{21}} & {s_{12}}&0\cr
-{s_{12}}& {c_{12}}&0\cr
0&0&1\cr} 
\label{eq:evol.2} 
\end{equation}
where $c_{ij}\equiv\cos\theta_{ij}$ and $s_{ij} \equiv \sin\theta_{ij}$.
Notice that, since the two Majorana phases do not affect neutrino 
oscillations, they are not included in the expression above.
The angles $\theta_{ij}$ 
can be taken without 
loss of generality to lie in the first quadrant, $\theta_{ij}\in[0,\pi/2]$.  
Concerning the CP violating phase $\delta$ we chose 
the convention $0\leq \delta \leq \pi$  and  two choices of mass ordering.

For solar and atmospheric oscillations, 
the required mass differences satisfy
\begin{equation}
\Delta m^2_\odot=\Delta m^2 \ll \Delta M^2=\Delta m^2_{\rm atm}.
\label{deltahier}
\end{equation}
so there are two possible mass orderings which, without any 
loss of generality can be chosen to be  as shown in Fig.~\ref{schemes}.
The direct scheme is naturally related to hierarchical masses,
$m_1\ll m_2\ll m_3$, for which $m_2\simeq\sqrt{\Delta m^2_{21}}$ and 
$m_3\simeq\sqrt{\Delta m^2_{32}}$
On the other hand, the inverted scheme implies that $m_3< m_1\simeq m_2$.
In both cases neutrinos can have 
quasi-degenerate masses, $m_1\simeq 
m_2\simeq m_3\gg \Delta m^2_{21}, |\Delta m^2_{32}|$. 
The two orderings are often referred to in terms of the
sign($\Delta m^2_{31}$).   
\begin{figure}[h]
\begin{center}
\includegraphics[scale=0.4]{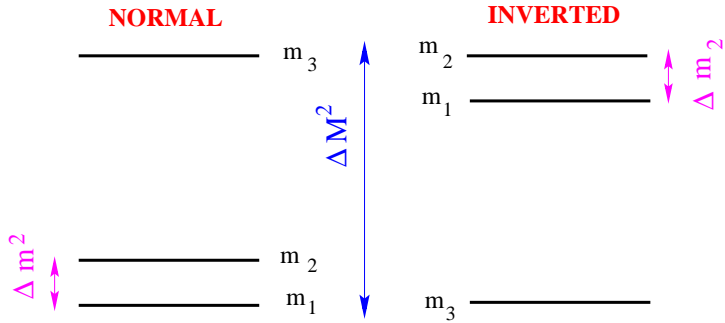}
\end{center}
\caption{Mass schemes for 3 $\nu$ oscillations}
\label{schemes}
\end{figure}

Thus, in total the three-neutrino oscillation analysis involves 
seven parameters: 2 mass
differences, 3 mixing angles, the CP phase and the sign($\Delta m^2_{31}$).
In general the transition probabilities present an oscillatory behaviour with 
two oscillation lengths. 
For transitions in vacuum, the results for normal and inverted schemes
are equivalent. In the presence of matter effects, this is 
no longer valid.
However, the hierarchy in the splittings, 
Eq.~({\ref{deltahier}), leads to important simplifications. \\
-- For solar neutrinos the oscillations with the atmospheric oscillation 
length are averaged out and the survival probability takes the form:
\begin{equation}
P^{3\nu}_{ee,MSW}
=\sin^4\theta_{13}+ \cos^4\theta_{13}P^{2\nu}_{ee,MSW} 
\label{p3}
\end{equation}
where $P^{2\nu}_{ee,MSW}$ 
is obtained with the modified sun density $N_{e}\rightarrow \cos^2\theta_{13} N_e $. 
So the analyses of solar data constrain
three of the seven parameters: 
$\Delta m^2_{21}, \theta_{12}$ and $\theta_{13}$. The effect of $\theta_{13}$ is
to decrease the energy dependence of the solar survival probability.
\\
-- For atmospheric neutrinos, 
the solar wavelength is too long
and the corresponding oscillating phase is negligible. As a consequence
the atmospheric
data analysis restricts $\Delta m^2_{31}\simeq \Delta m^2_{32}$, 
$\theta_{23}$ and
$\theta_{13}$, the latter being the only parameter common to both solar 
and atmospheric neutrino oscillations and
which may potentially allow for some mutual influence. The effect of
$\theta_{13}$ is to add a $\nu_\mu\rightarrow\nu_e$ contribution to the
atmospheric oscillations. \\
-- At reactor experiments the solar wavelength is unobservable if 
$\Delta m^2< 8\times 10^{-4}$ eV$^2$ and the relevant survival probability
oscillates with wavelength determined by $\Delta m^2_{31}$ and  
amplitude determined by $\theta_{13}$. 

CP is unobservable in the present data under the condition~(\ref{deltahier}). 
There is, in principle some
dependence on the normal versus inverted orderings due to matter effects
in the Earth for atmospheric neutrinos, controlled by the mixing angle
$\theta_{13}$.  
In Fig.~\ref{fig:atmos3} I show the allowed regions for the oscillation
parameters $\Delta M^2$ and $\cos\delta\sin^2\theta_{23}$ 
from the analysis of the 
atmospheric neutrino data in the framework of three-neutrino oscillations
~\cite{michele} for different values of $\sin^2\theta_{13}$. 
The figure illustrates how, under the condition (\ref{deltahier}), 
there is no dependence on the CP phase, $\delta$, while for large
values of $\theta_{13}$ the results for normal and inverted schemes
are different.  
\begin{figure}[h]
\begin{center}
\includegraphics[scale=0.33]{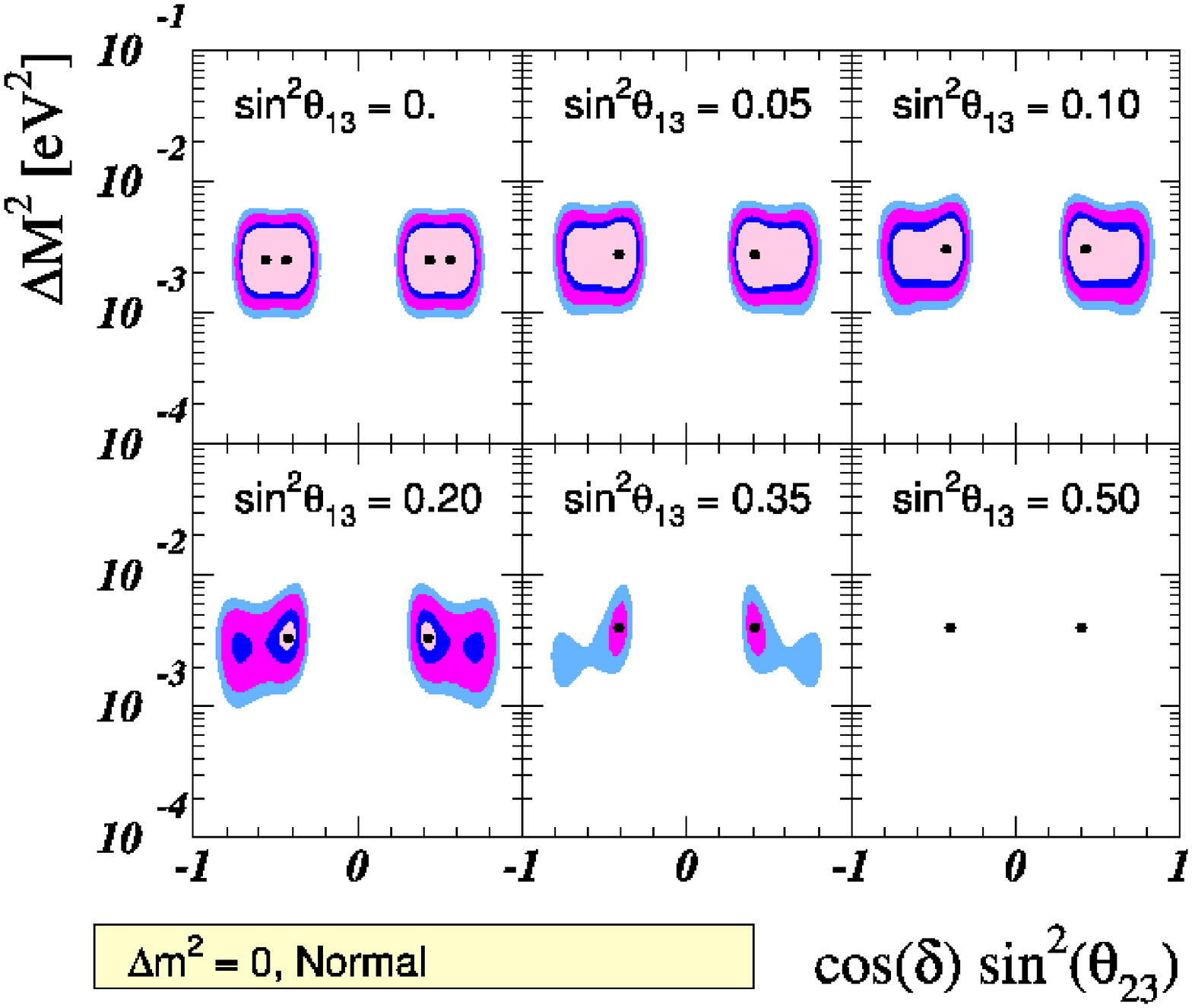}
\vskip -0.2cm
\includegraphics[scale=0.33]{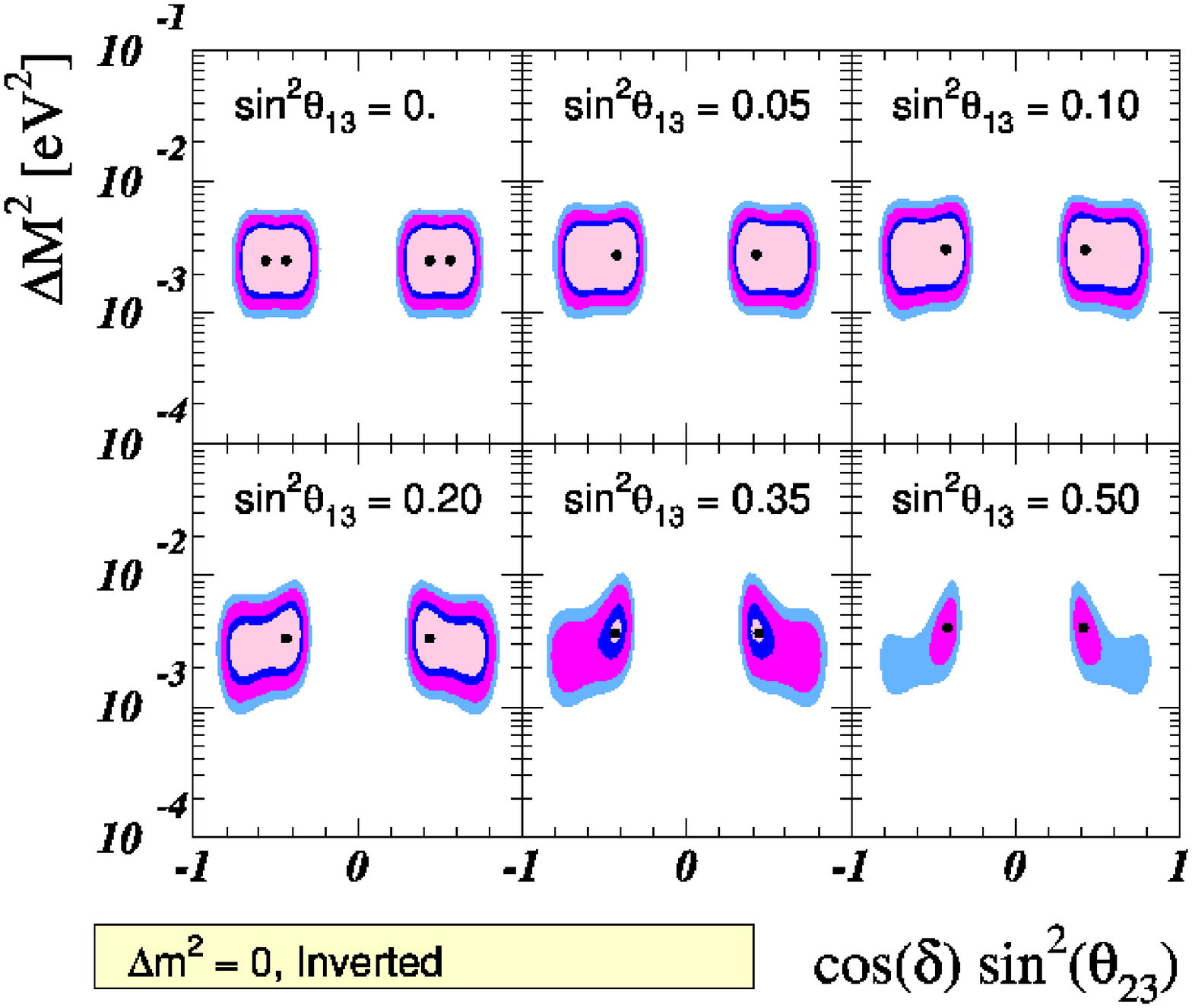}
\end{center}
\caption{
      90\%, 95\%, 99\% and $3\sigma$ allowed regions in the
      $(\cos^2\delta\sin^2\theta_{23}, 
      \Delta M^2)$ plane, for different values of
      $\sin^2\theta_{13}$ from the analysis of the
      atmospheric neutrino data.}
\label{fig:atmos3}
\end{figure}
From the analysis we also find that the atmospheric neutrino data favours
$\theta_{13}$ small and an upper bound on this angle is obtained.
This is due to the fact that the atmospheric data give no evidence 
for $\nu_e$ oscillation. 
 
Actually all data independently favours  $\theta_{13}=0$: the solar 
data exhibit energy dependence and this imposes a bound on  $\theta_{13}$
although not very strict. Most important, reactor data exclude
$\bar\nu_e$-disappearance's at the atmospheric wavelength. 
In Fig.~\ref{teta13} I plot the results of the analysis 
of solar, atmospheric and reactor data on the allowed values of $\theta_{13}$.
\begin{figure}
\begin{center}
\includegraphics[scale=0.45]{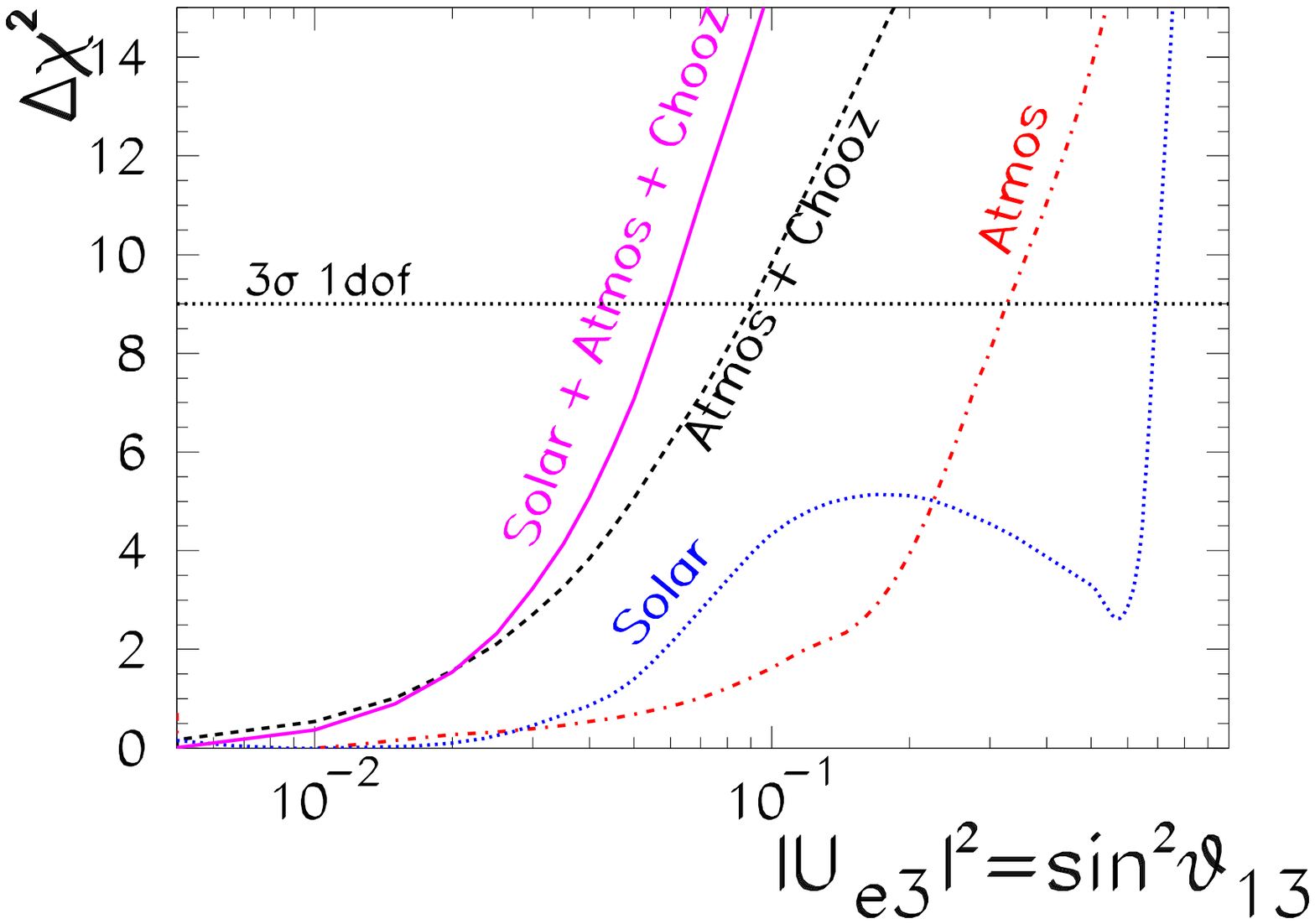}
\end{center}
\caption{Dependence of $\Delta\chi^2$ on $\sin^2\theta_{13}$ in the analysis
of the atmospheric, solar and CHOOZ neutrino data. The dotted horizontal line 
corresponds to the 3$\sigma$ limit for a single parameter.}
\label{teta13}
\end{figure}
The combined analysis results in a limit
$\sin^2\theta_{13}\leq 0.06$ at 3$\sigma$~\cite{michele,review}. 
Within this limit the difference between normal and inverted orderings in 
atmospheric neutrino data is  below present experimental sensitivity.  
Thus, under the condition (\ref{deltahier}) the three-neutrino analysis
of solar+atmospheric+reactor data effectively involves only five parameters.

The global analysis of solar, atmospheric and reactor data 
in the five-dimensional parameter space leads to the following allowed 
3$\sigma$ ranges for individual 
parameters (that is, when the other four parameters have been chosen to 
minimize the global $\chi^2$):
\begin{eqnarray}
2.4\times 10^{-5}<& \Delta m^2_{21}/\mbox{\rm eV$^2$}&<
2.4\times 10^{-4}\;\;\;\;\;{\rm LMA}\; 
\nonumber \\
0.27<&\tan^2\theta_{12}&< 0.77 \;\;\;\;\;\;\;\;\;\;\;\;\;\;\;
{\rm LMA} \;
\nonumber \\
1.4\times 10^{-3}<& \Delta m^2_{32}/\mbox{\rm eV$^2$}&< 
6.0\times 10^{-3} 
\nonumber\\ 
0.4<&\tan^2\theta_{23}&< 3.0, 
\nonumber\\ 
\sin^2\theta_{13} &< 0.06 & 
\label{globalranges}
\end{eqnarray} 
These results can be translated into our present knowledge of the
moduli of the mixing matrix $U$:
\begin{equation}
|U| =\pmatrix{ 0.73-0.89&0.45-0.66&<0.24 \cr
0.23-0.66&0.24-0.75&0.52-0.87\cr 
0.06-0.57&0.40-0.82&0.48-0.85\cr}.   
\end{equation}
\subsection{The Future: Long Baseline Experiments}
$\nu_\mu$  oscillations with  $\Delta m^2_{\rm atm}$ are being probed
and will be further tested  using accelerator beams at 
LBL experiments. In these experiments the intense neutrino  
beam from an accelerator is aimed at a detector located underground at a 
distance of several hundred kilometers. 
At present there are three such projects approved: K2K~\cite{k2k} 
which runs with a baseline of about 
235 km from KEK to SK, MINOS~\cite{minos} 
under construction with a baseline of 730 km from Fermilab to 
the Soudan mine where the detector will be placed, and 
two detectors, OPERA~\cite{opera} and ICARUS~\cite{icarus}  
under construction with a baseline of 730 km from CERN to Gran Sasso.
\begin{figure}[ht]
\begin{center}
\includegraphics[scale=0.55]{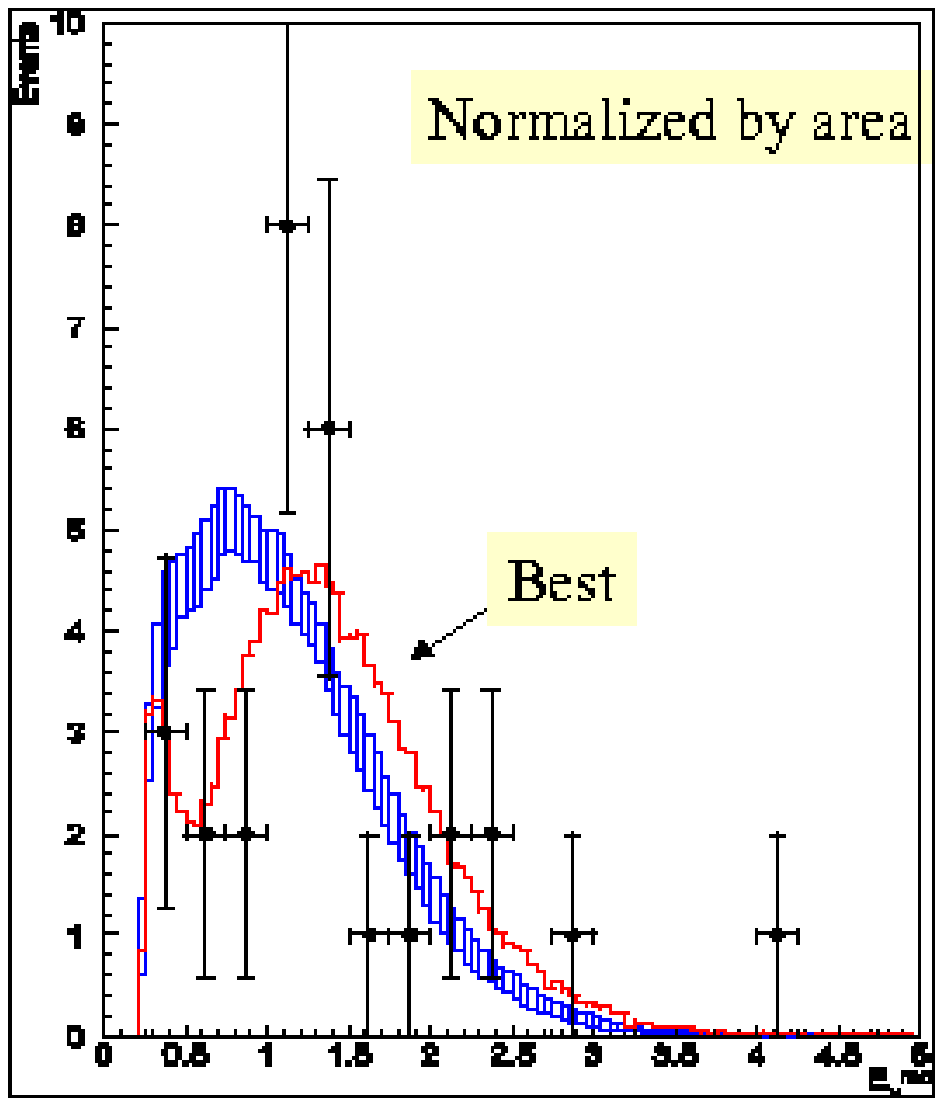}
\includegraphics[scale=0.45]{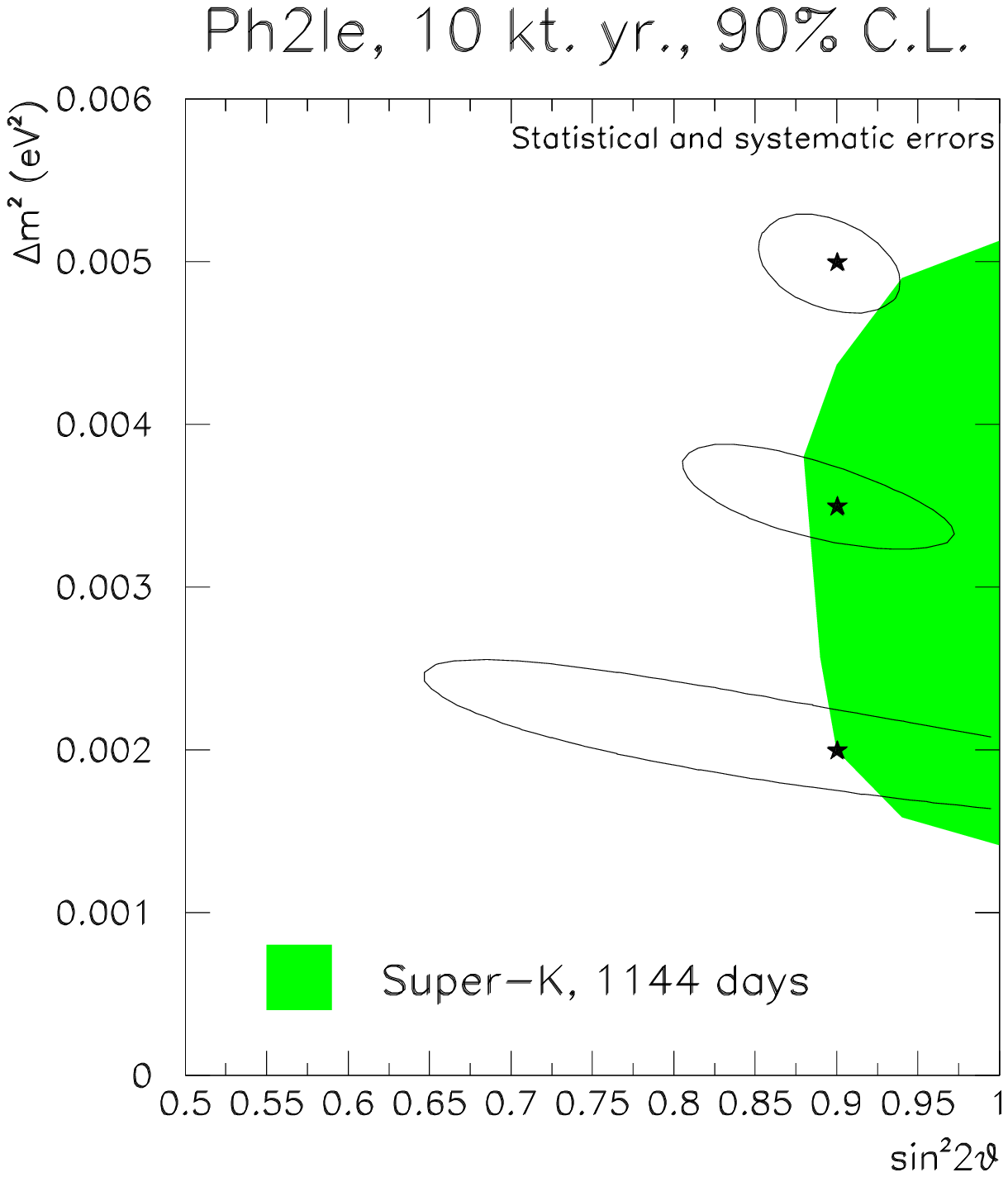}
\end{center}
\caption{{\sl Left}: Observed events at the K2K experiment, together
with the expected distribution for the case of nooscillation and 
oscillations with the best fit point of the atmospheric neutrino analysis.
{\sl Right:} Expected reconstruction of the oscillation parameters in 
Minos.}
\label{fig:minos}
\end{figure}

The first results from K2K seem to confirm the atmospheric 
oscillations (see Fig~\ref{fig:minos}) but statistically they 
are still not very significant. In the near future K2K 
will accumulate more data enabling it to confirm the
atmospheric neutrino oscillation. Furthermore, combining the
K2K and atmospheric neutrino data will lead to 
a better determination of the mass and mixing parameters.

In a longer time scale, the results from MINOS will provide more 
accurate determination of these parameters as shown in Fig.~\ref{fig:minos}. 
OPERA and ICARUS are designed to observe the $\nu_\tau$ appearance. 
MINOS, OPERA and ICARUS have certain sensitivity to $\theta_{13}$ although
by how much they will be ultimately able to improve the present 
bound is still undetermined.
\section{LSND and Sterile Neutrinos}
\subsection{The Evidence}
The only positive signature of oscillations at a laboratory experiment comes 
from the Liquid Scintillator Neutrino Detector (LSND) which run at Los Alamos 
Meson Physics Facility.
The primary neutrino flux comes from $\pi^+$'s produced in a 30-cm-long water
target when hit by protons from the LAMPF linac with  800 MeV kinetic energy. 
Most of the produced $\pi^+$'s come to rest and decay through the sequence
$\pi^+\to\mu^+\nu_{\mu}$, followed by $\mu^+\to e^+\nu_e\bar\nu_\mu$. The
$\bar\nu_\mu$'s so produced have a maximum energy of $52.8$ MeV. 
This is called
the {\it decay at rest} (DAR) flux and is used to study 
$\bar\nu_\mu\to \bar\nu_e$ oscillations. 
For DAR related measurements, $\bar\nu_e$'s are detected in the quasi 
elastic process $\bar\nu_e\,p\to e^{+}\,n$, in correlation with a 
monochromatic 
photon of $2.2$  MeV arising from the neutron capture reaction 
$np\to d\gamma$. Their final result is an excess of events 
of $87.9\pm22.4\pm6$ 
events, corresponding to an oscillation probability of 
$(2.64\pm0.67\pm0.45)\times10^{-3}$. 
\subsection{The Interpretation: Four Neutrino Mixing}
In the two-family formalism the LSND results 
lead to the oscillation parameters shown in Fig.~\ref{fig:miniboone}. 
The shaded 
regions are the 90~\% and 99~\% likelihood regions from LSND. The best fit 
point corresponds to $\Delta m^2=1.2$ eV$^2$ and $\sin^22\theta=0.003$.

The region of parameter space which is favoured by the LSND observations
has been partly tested by other experiments in particularly by the 
KARMEN experiment\cite{karmen}. The KARMEN experiment was performed at 
the neutron spallation facility ISIS of the Rutherford Appleton Laboratory. 
They found no evidence of flavour transition which translated into 
exclusion curve in the two-neutrino parameter space given in 
Fig.~\ref{fig:miniboone} 
\begin{figure}[ht]
\vglue -0.5cm
\begin{center}
\includegraphics[scale=0.4]{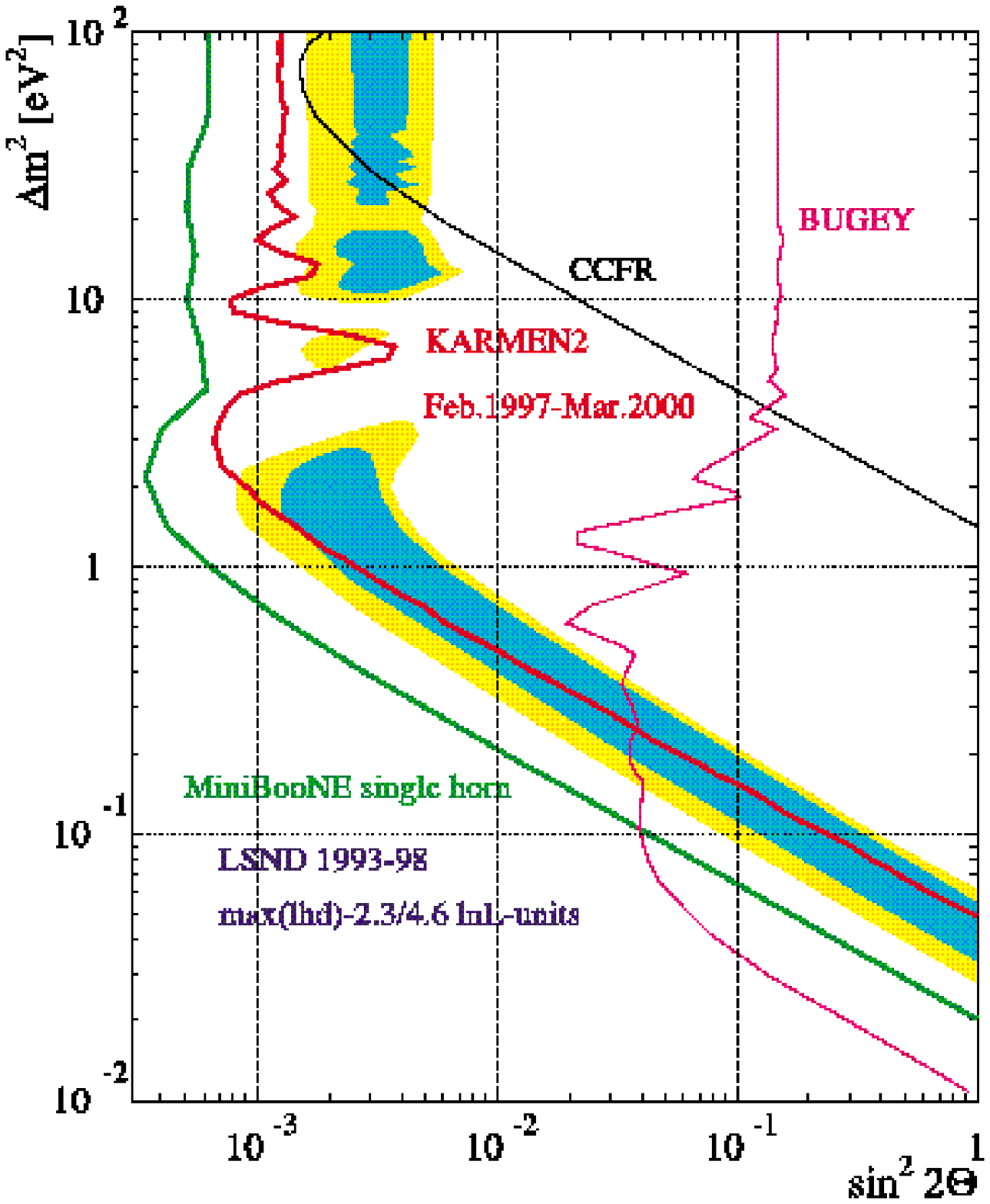}
\includegraphics[scale=0.7]{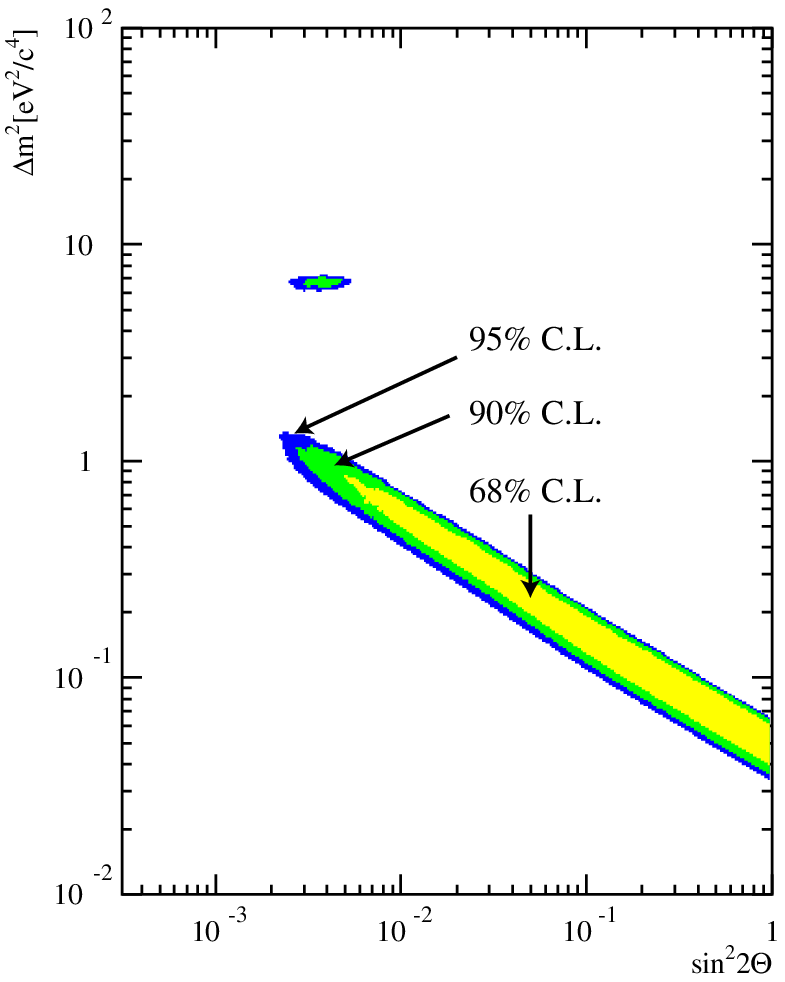}
\end{center}
\caption{{\sl Left:}
Allowed regions (at 90 and 99 \% CL) for $\bar\nu_e\to\bar\nu_\mu$ 
oscillations from the LSND experiment compared with the exclusion regions 
(at 90\% CL) from KARMEN2 and other experiments. The 90 \% CL expected 
sensitivity curve for MiniBooNE is also shown.
{\sl Right:} Allowed region from the combined analysis of LSND and
KARMEN2 data.}
\label{fig:miniboone}
\end{figure}

Recently a combined analysis of the LSND and KARMEN data has been 
performed~\cite{lsndkarmen} which shows that both results can 
still be compatible within the parameter region shown in the 
left panel of Fig.~\ref{fig:miniboone}.

To accommodate the LSND result together with the solar and atmospheric
data in a single neutrino oscillation framework, there must be at least 
three different scales of neutrino mass-squared differences which
requires the existence of a fourth light neutrino.
The measurement of the decay width of the
$Z^0$ boson into neutrinos makes the existence of three, and only three, light 
active neutrinos an experimental fact. Therefore, the fourth neutrino 
must not couple to the standard electroweak current, that is, 
it must be sterile.

One of the most important issues in the context of four-neutrino scenarios is 
the four-neutrino mass spectrum. There are six possible four-neutrino schemes,
shown in Fig.~\ref{fig:4mass}, that can accommodate the results from solar and 
atmospheric neutrino experiments as well as the LSND result. They can be 
divided in two classes: (3+1) and (2+2).
In the (3+1) schemes, there is a group of three close-by neutrino masses that 
is separated from the fourth one by a gap of the order of 1~eV$^2$, which is 
responsible for the SBL oscillations observed in the LSND experiment. 
In (2+2) schemes, there are two pairs of close masses separated by the LSND 
gap. The main difference between these two classes is the following: if a
(2+2)-spectrum is realized in nature, the transition into the sterile neutrino
is a solution of either the solar or the atmospheric neutrino problem, 
or the sterile neutrino takes part in both, whereas with a (3+1)-spectrum the
sterile neutrino could be only slightly mixed with the active
ones and mainly provide a description of the LSND result.
\begin{figure}
\begin{center}
\includegraphics[scale=0.8]{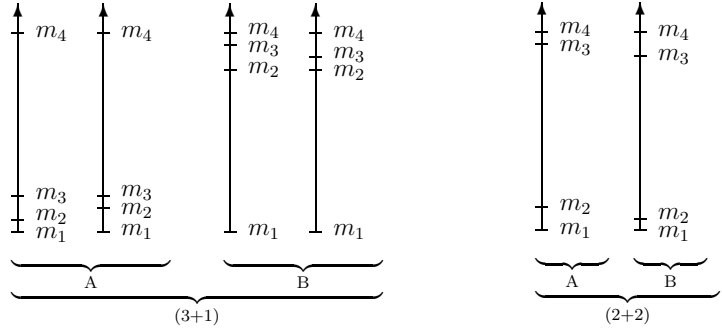}
\end{center}
\caption{The six types of 4-neutrino mass spectra.} 
\label{fig:4mass}
\end{figure}

The phenomenological situation at present is that none of the four-neutrino
scenarios are favoured by the data \cite{valle}. In brief 
(3+1)-spectra are disfavoured by the incompatibility between the LSND
signal and the present constraints
from short baseline laboratory experiments, while (2+2)-spectra are disfavoured by
the existing constraints from the sterile oscillations in solar and
atmospheric data. 
\begin{figure}[ht]
\begin{center}
\mbox{\epsfig{file=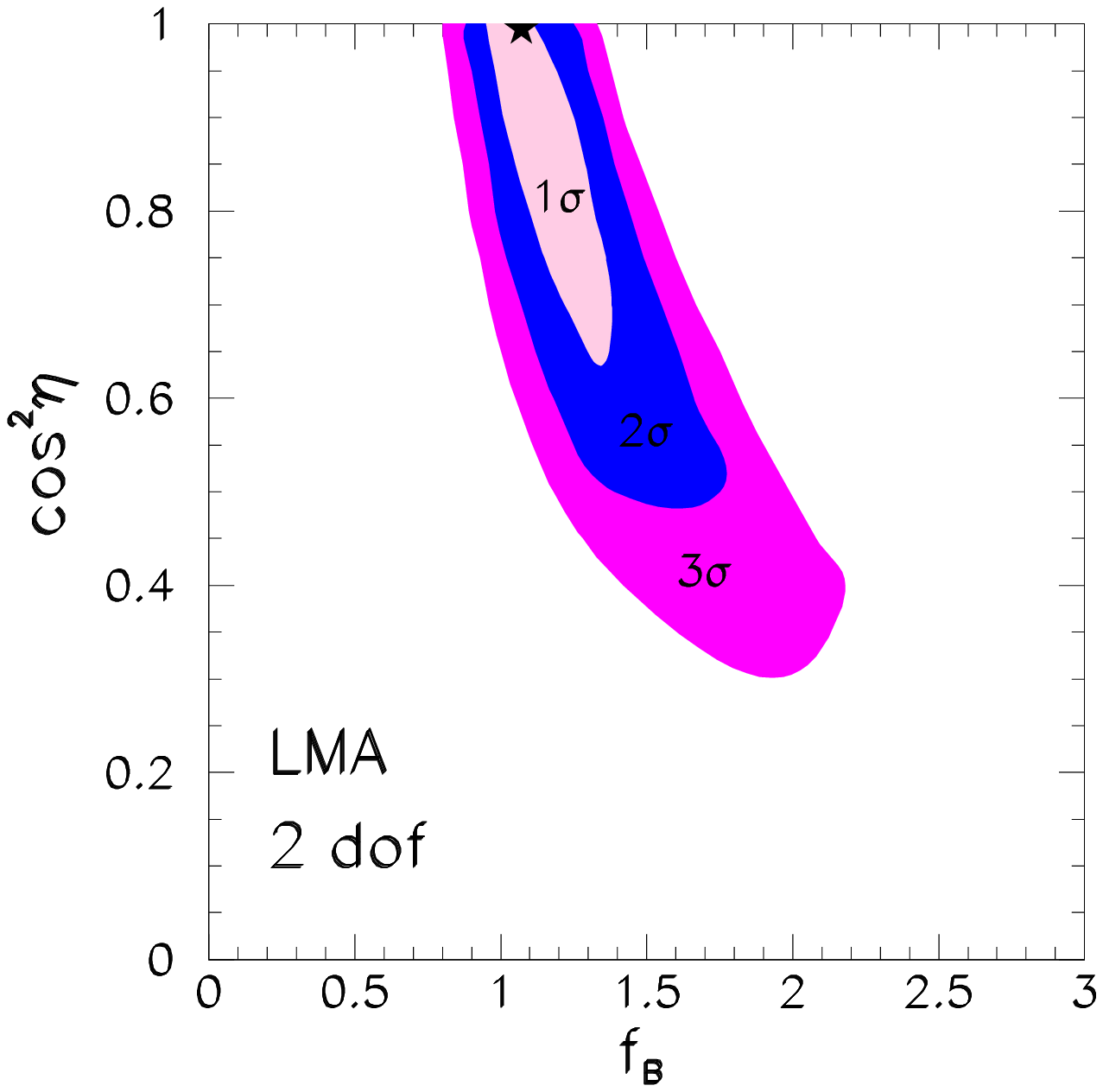,width=0.45\textwidth,height=0.3 \textwidth} }
\mbox{\epsfig{file=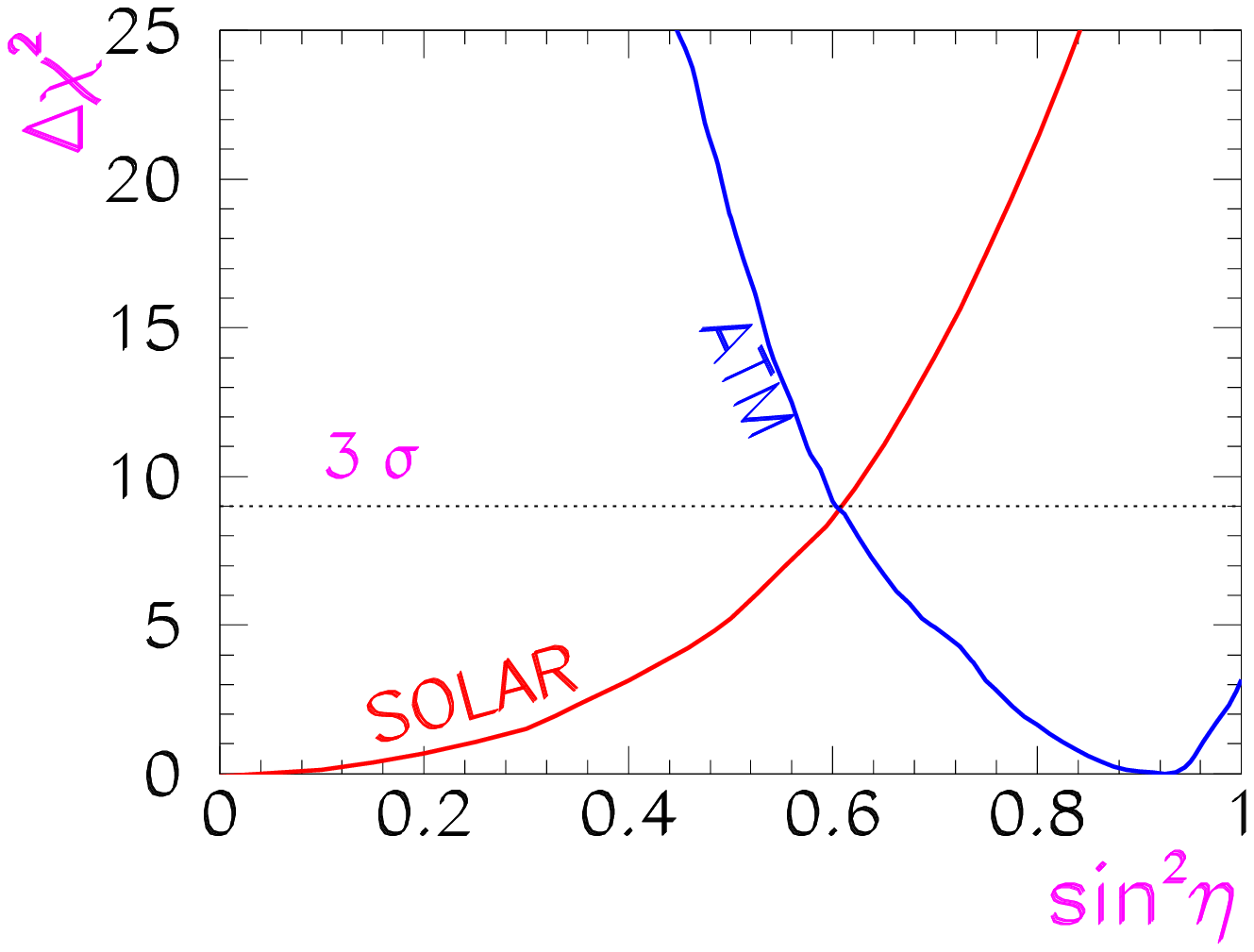,width=0.45
\textwidth,height=0.3 \textwidth} }
\end{center}
\caption{Left: Constraint on the active-sterile admixture 
in solar neutrino oscillations versus the $^8$B neutrino flux 
enhancement factor. 
Right: Present status of the bounds on the active-sterile admixture from
solar and atmospheric neutrino data in (2+2)-models.
}
\label{ster}
\end{figure}

In this respect it has been recently pointed out that
the existing constraint on the sterile admixture in the solar neutrino
oscillations can be relaxed if the $^8$B neutrino flux is allowed to be
larger than in the SSM by a factor $f_B$~\cite{barger,sterile}. 
The analysis  is performed in
the context of  solar conversion  $\nu_e
\rightarrow \nu_x$, where $\nu_x = \cos \eta~ \nu_a + \sin \eta ~\nu_s$.
In Fig.~\ref{ster} I show the presently allowed range of $\eta$ 
as a function of $f_B$. The obtained upper bound on $\sin^2\eta$ from this
most general solar analysis has to be compared with the corresponding lower
bound from the analysis of atmospheric data. In Fig.~\ref{ster} I show the
corresponding comparison (the curve for the atmospheric data is taken from 
Ref.~\cite{valle}). 
\subsection{The Future: MiniBooNE}
The MiniBooNE experiment~\cite{miniboone}, 
will be able to confirm 
or disprove the LSND oscillation signal within the next two years 
(see Fig.~\ref{fig:miniboone}). 
Should the oscillation signal be confirmed as well as the solar 
signal in KamLAND and  the atmospheric in LBL experiments, we will 
face the challenging 
situation of not having a successful  ``minimal'' phenomenological 
description at low-energy of the leptonic mixing. 

Alternative explanations to the LSND observation include the 
possibility of
CPT violation~\cite{CPT} which would imply that the mass differences and
mixing  among neutrinos would be different from the ones for antineutrinos.
This scenario is being tested with the running experiments. 
An imminent test of CPT will be the comparison of 
the observation in KamLAND of $\overline{\nu_e}$ disappearance versus 
solar $\nu_e$ disappearance. Also, at present, MiniBooNE~\cite{miniboone}  
is running in the neutrino mode  searching for $\nu_\mu\rightarrow\nu_e$ to
be compared with the antineutrino signal in LSND. Thus  
an oscillation  signal at KamLAND or MiniBooNE will
put serious constraints on CPT violation for $\nu$'s. 

\section{Neutrino Mass Scale}
\label{direcdet}
Oscillation experiments provide information on
$\Delta m^2_{ij}$, 
and on the leptonic mixing angles, $U_{ij}$. But they are insensitive to 
the absolute mass scale for the neutrinos. 
Of course, the results of an oscillation experiment do provide a lower bound
on the heavier mass in $\Delta m^2_{ij}$, $|m_i|\geq\sqrt{\Delta m^2_{ij}}$ for
$\Delta m^2_{ij}>0$. But there is no upper bound on this mass. In particular,
the corresponding neutrinos could be approximately degenerate at a mass
scale that is much higher than $\sqrt{\Delta m^2_{ij}}$. 
Moreover, there
is neither upper nor lower bound on the lighter mass $m_j$.

Information on the neutrino masses, rather than mass differences, can be 
extracted from kinematic studies of reactions in which a neutrino or an 
anti-neutrino is involved. In the presence of mixing the most 
relevant constraint comes from Tritium beta decay 
${\rm ^3H \rightarrow\ ^3He + e^-+\overline\nu_e}$ 
which, within the present and expected
experimental accuracy, can limit the combination 
\begin{equation}
m_{\beta}=\sum_i m_i |U_{ei}|^2
\label{mbeta}
\end{equation}
The present bound is 
$m_{\beta}\leq 2.2$ eV at 95 \% CL~\cite{beta}.
A new experimental project, KATRIN, is under consideration with an estimated 
sensitivity limit: $m_{\beta}\sim0.3$ eV. 

Direct information on neutrino masses can also 
be obtained from neutrinoless double beta decay
$(A,Z) \rightarrow (A,Z+2) + e^{-} + e^{-}$.
The rate of this process is proportional to the 
{\it effective Majorana mass of $\nu_e$},
\begin{equation}
m_{ee}=\left| \ \ \sum_i m_i U_{ei}^2 \ \ \right|
\end{equation}
which, unlike Eq.~(\ref{mbeta}),
depends also on the three CP violating phases. 
Notice that in order to induce the $2\beta0\nu$ decay, $\nu$'s must 
Majorana particles. 

The present strongest bound from $2\beta0\nu$-decay is 
$ m_{ee} < 0.34$ eV at 90 \% CL~\cite{klapdor}. 
Taking into account systematic errors related to nuclear matrix elements,
the bound may be weaker by a factor of about 3. A sensitivity of 
$m_{ee}\sim 0.1$ eV is expected to be reached by the currently running 
NEMO3 experiment, while a series of new 
experiments is planned with sensitivity of up to  
$m_{ee} \sim 0.01$ eV~\cite{vogel} (see Table~\ref{betabeta} for some of these 
proposals). 
\begin{table}
\begin{center}
\begin{tabular}{|c|c|c|c|}
Experiment & Search& Source  & goal \\
\hline 
KATRIN & {$^3$H $\beta$-dec} &   & ${m_\beta}<0.3$-- 0.4 eV\\ 
\hline
{ NEMO3}      & { $(\beta\beta)_{0\nu}$} & $^{100}$Mo &  
{ $|\langle m_{ee}\rangle|\sim 0.2$ eV} \\
CUORE      & { $(\beta\beta)_{0\nu}$} & $^{130}$Te  &  
$|\langle m_{ee}\rangle|\sim 0.03$ eV \\
EXO      & { $(\beta\beta)_{0\nu}$} & $^{136}$Xe  &  
$|\langle m_{ee}\rangle|\sim 0.05$ eV \\
MOON    & { $(\beta\beta)_{0\nu}$} &   $^{100}$Mo & 
$|\langle m_{ee}\rangle|\sim 0.04$ eV \\
GENIUS      & { $(\beta\beta)_{0\nu}$} & $^{76}$Ge&  
$|\langle m_{ee}\rangle|\sim 0.01$ eV \\
Majorana  & { $(\beta\beta)_{0\nu}$} & $^{76}$Ge&  
$|\langle m_{ee}\rangle|\sim 0.02$ eV 
\end{tabular}
\end{center}
\label{betabeta}
\caption{Future experiments and their proposed sensitivity 
to the relevant neutrino mass scale.} 
\end{table}

The knowledge of $m_{ee}$ can provide information on the mass
and mixing parameters that is independent of the $\Delta m^2_{ij}$'s. However,
to infer the values of neutrino masses, additional assumptions are required.
In particular, the mixing elements are complex and may lead to strong 
cancellation, $m_{ee}\ll m_1$. Yet, the combination of results
from $2\beta0\nu$ decays and Tritium beta decay can test and,
in some cases, determine the mass parameters of given
schemes of neutrino masses~\cite{bb} provided that the nuclear matrix elements
are known to good enough precision. 
%
%\vspace*{-1cm}
%\section*{Acknowledgements}
\vskip 0.3cm
{\sl This work is supported by 
the MC fellowship HPMF-CT-2000-00516 and by the Spanish
DGICYT grant FPA2001-3031.} 

\end{document}